# Low-Cost Lung Cancer Detection Using Machine Learning on Breath Samples

Jayanth R. Mokkapati

# Table of Contents

## Introduction

Lung cancer is the leading cause of cancer-related death in this generation and is predicted to stay that way for the foreseeable future. Early diagnosis and treatment of lung cancer symptoms can stop the cancer cells from spreading throughout the body. It is possible to create a sustainable prototype model for the treatment of lung cancer using current advancements in computational intelligence without harming the environment. It will reduce the number of resources squandered and the amount of labor needed to complete manual activities, which will save time and money. The creation of the Al Nose was the best way to address the issue of cancer-related fatalities because every family could purchase and utilize one.

Electronic nose devices, such as the AI Nose, have gained popularity recently as a method for detecting lung cancer and other respiratory diseases. The concepts and patterns of sensor responses in conventional e-nose systems have been highly constant. Fifty breath samples were examined using a miniature e-nose device that has 14 gas sensors of four different sorts. The effectiveness of the system in differentiating lung cancer from other respiratory illnesses and healthy controls was evaluated using five feature extraction approaches, two classifiers, and healthy controls. Finally, it examined how different sensor kinds affect e-nose systems' ability to recognize things. Using the PLS-DA (Partial Least Squares Discrimination Analysis) classification technique, the sensitivity, specificity, and accuracy of differentiating lung cancer patients from healthy controls are 91.378 percent, 92.432 percent, and 93.587 percent, respectively.

Additionally, a machine learning model based on support vector machines (SVMs) was employed to enhance the identification procedure using the lung cancer dataset. Lung cancer patients were categorized according to their symptoms using an SVM classifier, and the Python programming language was also used to advance the model's implementation. To evaluate the effectiveness of the SVM model, a number of criteria were taken into consideration. The assessed model was put to the test using a variety of cancer datasets from the UT Southwestern Medical Center. Communities will be able to give their inhabitants better healthcare as a result of the favorable outcomes of this study. Lung cancer patients can obtain real-time therapy with the least amount of effort and delay, at any time and from any location. The proposed model and the current SVM and SMOTE algorithms were contrasted.

The results suggest that type-specific sensors may soon significantly increase the diagnostic accuracy of e-nose devices. For the most accurate results, this gadget uses the element of machine learning with numerous models like PLS-DA and SVM's, while taking into consideration all the many types of sensors it has. In conclusion, the AI Nose developed in this study achieved an accuracy of 98.2 percent after several testing, but further research is being done in the hopes that it may eventually be available for purchase by the general public. The effects that cutting-edge technology, like AI Noses, may have on our societies are unlimited if this area continues to grow and machine learning is used.

## Materials, Methodology, and Procedures

A wide range of sectors, including agricultural, aviation, cosmetics, food and beverage, healthcare, and military applications, employ electronic nose technology. These gadgets use a variety of sensors, a database, and a pattern recognition algorithm to detect and identify scents, simulating the human olfactory system in the process. The brain detects an odor when someone breathes through their nose by observing the chemical interactions between volatile substances in the nasal cavity and olfactory receptors. Similar to chemical sensors, electronic noses produce electrical impulses when they respond to volatile substances. These signals are then captured and utilized to build a computerized representation of various odors that is simple to recall and contrast.

However, it is important to note that electronic noses are not as sensitive or selective as the human nose. A sensor array, a data processing system, and a pattern recognition system make up the electronic nose system. The electrical responses of the sensors change when chemicals are added to the system depending on the characteristics of the molecules. Depending on the substances they come into touch with, the sensors' conductivity also varies. Organic and inorganic volatile substances, as well as nitrogen, oxygen, and carbon dioxide, are all present in human exhaled breath. Since the amounts of these volatile organic compounds (VOCs) in the breath of healthy people and patients with certain disorders might vary, these VOCs can be utilized as biomarkers for specific diseases. Based on the quantity of the molecules that will be covered in this paper, VOCs serve as the identifying factor to assess whether a patient has lung cancer.

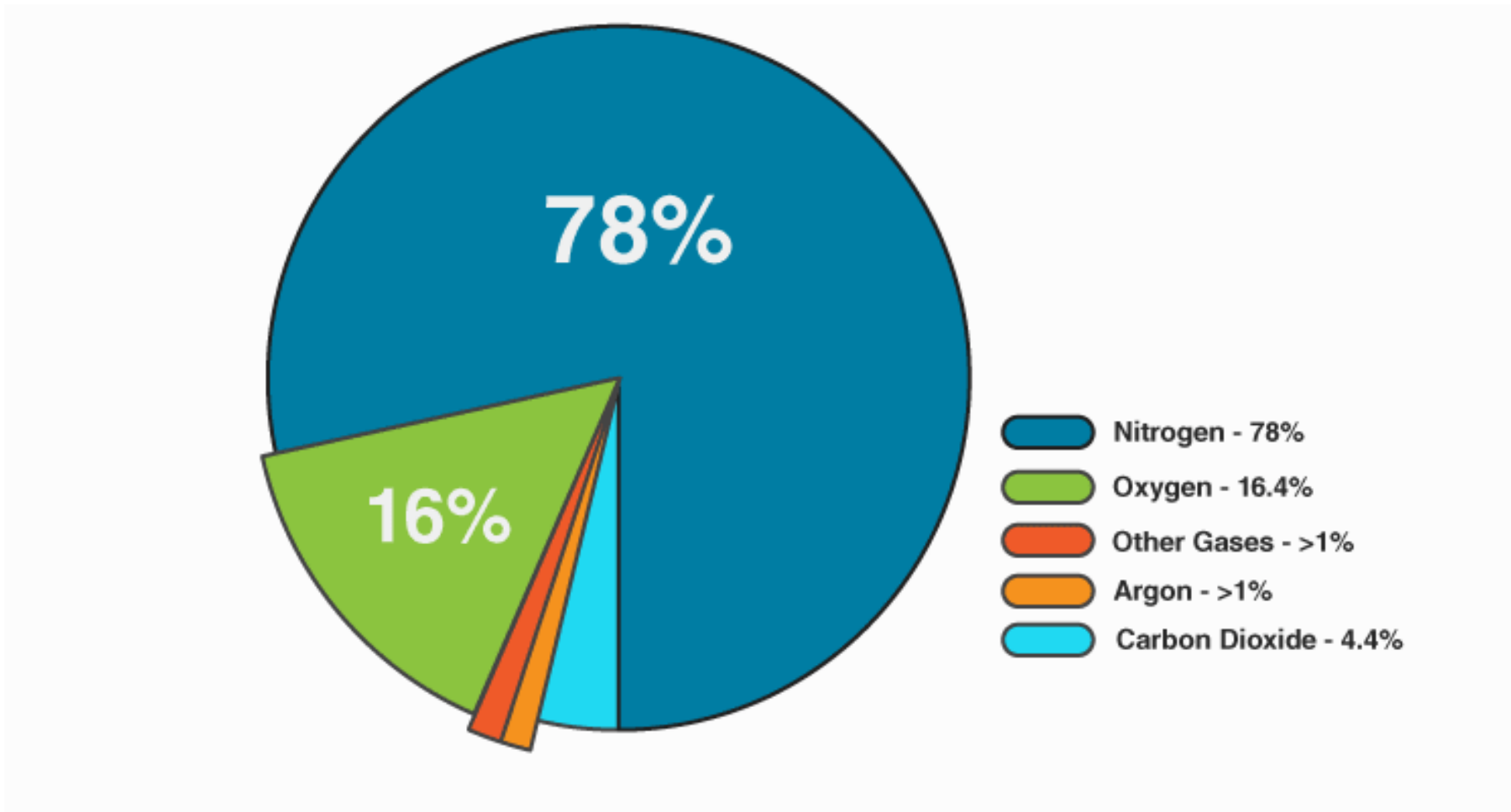

**Figure 1. Components of Exhaled Breath Utilized By AI Nose**

Electronic noses made up of several gas sensor arrays that analyze breath have emerged as a viable method for identifying lung disorders. Electronic noses built on metal oxide semiconductors are particularly well-liked for evaluating exhaled breath to find VOCs. Methods for analyzing breath, including gas chromatography-mass spectrometry, ion mobility spectrometry, and gas chromatography-flame ionization detection, are costly, time-consuming, and labor-intensive to do. Electronic nose analysis, on the other hand, is an easy, quick, and affordable technique that can offer improved accuracy, sensitivity, and specificity for diagnosing illnesses like lung cancer.

The goal of the study was to develop an electronic nose (e-nose) device powered by artificial intelligence (AI) that could distinguish between exhaled breath samples from lung cancer patients and those from healthy people. Metal oxide semiconductor (MOS) sensors from Figaro USA (TGS sensors) were used to construct the e-nose system, and data was gathered using an Arduino Uno developer board. After then, Python machine learning methods were used to examine the data. To verify its efficacy, the system was put to the test on a sample of 24 lung cancer patients and 40 healthy controls.

## What Are Biomarkers?

By definition, a biomarker is "a characteristic that is measured as an indicator of normal biological processes, pathogenic processes, or responses to an exposure or intervention, including therapeutic interventions," according to the Biomarker Working Group of the US Food and Drug Administration and National Institutes of Health. Molecular, histologic, radiographic, or physiological characteristics are all possible for biomarkers. A biomarker does not serve as a measurement of a person's emotions, behaviors, or capacity for survival. Biomarker categories [5] include:

Diagnostic biomarkers, monitoring biomarkers, prognostic biomarkers, predictive biomarkers, susceptibility/risk biomarkers, pharmacodynamic/response biomarkers, and safety biomarkers. A biomarker has to fulfill specific criteria, such as being identified in diagnostic or screening situations at a time when it may change the patient's path through the disease or its outcome.

## Why Is There A Need For Biomarkers in Lung Cancer Detection?

### Detection biomarkers

A biomarker may be useful in three critical circumstances if it is available. When used with a thoracic cancer screening program that has a high risk of false-positive outcomes, it may first aid in the detection of lung cancer. Additionally, it could aid in the selection of high-risk individuals for a CT screening program. Such biomarkers have drawn interest in other fields (such as prostate-specific antigen and carcinoembryonic antigen) even though they have not been authorized for clinical use in lung cancer. Studies called ECLS and bio-MILD are now being conducted to assess blood tumour indicators for lung cancer. The gold standard for the diagnosis of thoracic malignancies is a pathology report with evidence of malignant cells. Sometimes, though, it is not possible to gather enough tissue to make a diagnosis (for instance, in patients with severe COPD who cannot be safely subjected to a biopsy), thus the multidisciplinary team makes its decisions on radiological and clinical data. These teams might benefit from a diagnostic biomarker in determining the risk of cancer.

### Predictive biomarkers

When forecasting a patient's reaction to treatment, a biomarker is essential. Biomarkers are often used to forecast therapy response in non-small cell lung cancer (NSCLC), such as the identification of epidermal growth factor receptor (EGFR) mutations, ALK mutations, and other driver mutations. The tumor percentage score for programmed death ligand 1 (PD-L1) is another biomarker utilized in NSCLC. Checkpoint inhibitor efficacy in stage IV and perhaps stage III NSCLC is correlated with PD-L1. Clinical studies for mesothelioma, small cell lung cancer, and various stages of NSCLC are also drawing attention to PD-L1. Due to the limited sensitivity and specificity of PD-L1 tests, there is now a lot of research being done to create new biomarkers or enhance existing ones, including PD-L1 to increase its sensitivity and specificity.

**What Are VOC's?**

Since the discovery of more than 3000 distinct VOCs in human breath in the 1970s, the discipline of breath analysis has swiftly expanded. VOCs, or volatile organic compounds, were initially detected. However, owing to metabolic activities within the human body, VOCs can also be exhaled or VOC patterns can be changed. The majority of particles in the air are created by external processes such as the environment and pollution. The idea that VOCs provide a distinct "smell" or "breathprint" for various diseases is based on the fact that these processes can be normal or can be produced by or affected by disease.

**VOCs In Other Devices**

It has been demonstrated that e-noses, which employ VOC detection and breathprint analysis, are useful for both the detection and diagnosis of non-respiratory illnesses including Barrett's esophagus and inflammatory bowel disease as well as respiratory illnesses like COPD and asthma. Positive e-nose research has also been reported for a number of cancer types, including bladder, colon, and head and neck cancer.

**VOCs In Lung Cancer**

Research has also been done on lung cancer, with early tests showing that trained dogs may identify lung cancer patients. Later, specific VOCs were found in lung cancer patients, but these techniques were found to be too costly, time-consuming, and inaccurate. Interest in VOCs for lung cancer diagnosis reemerged as technology advanced, and pattern recognition techniques were put into practice. These techniques "learn" patterns using artificial intelligence to distinguish between COPD or healthy participants and lung cancer patients. Widespread clinical usage of e-noses is now hindered by challenges with VOC stability and detection equipment, but if these problems can be overcome, it is envisaged that e-noses may become standard in practice. Additionally, progress has been made in mesothelioma detection using VOC analysis and breath analysis.

**E- Noses**

There are many types of e-noses, and they differ in how they sample the air. While some e-noses utilize a sample balloon to collect air, others use a canister. To prevent contamination and the effect of outside elements like disinfectants or cigarette smoke, all e-noses must be in a controlled environment.

Older e-noses may have measured individual VOCs, however more current e-noses analyze patterns using artificial intelligence, training on test sets, and validation using independent sets to give relevant data. The type of sensors employed varies significantly amongst e-noses; the most used techniques are gas chromatography, spectrometry, colorimetry, surface acoustic waves, and conductometry.

## Results and Findings

30 participants took part in the experiment. Twenty of them had been identified as having lung cancer, while the remaining eight had other lung problems despite having cancer-free tests. For each participant, air samples were taken from various locations along the respiratory system. In-situ samples from both lungs were taken using a modified bronchoscope. Before the bronchoscopic examination, each person's breath was also sampled, with the deeper alveolar breath being separated from the top part of the breath (known as the "dead space"). The second part of the breath, which is more appropriately known as mixed-expired breath, may still contain molecules produced in the upper airways since this separation does not entirely eradicate the effects of the dead space.

All samples save the dead space were subjected to electronic nose analysis; the samples were split between GC-MS and electronic nose analysis. All four types of samples were analyzed by GC-MS, but only for cancer patients. On electronic nose data for mixed-expired breath with a three latent variable model, partial least squares discriminant analysis (PLS-DA) was performed with a 93% accuracy rate. A leave-one-out method was used to calculate the quantity of latent variables in order to reduce prediction error. Due to the ambiguity of the data set, the PLS-DA models have not been confirmed by an independent dataset.However, cross-validated PLS-DA has been shown to be similar to a multivariate ANOVA statistical test with the extra advantage of score plots that enable the visualization of the link between the variables. Fig. 2's depiction of the first two latent variables reveals how the electronic nose data clusters. This outcome is in line with other discoveries made with the same sensor technology and electronic noses. In particular, it supports the notion that, when the same breath sampling protocol was employed, cancer results tend to be less consistent than control data.

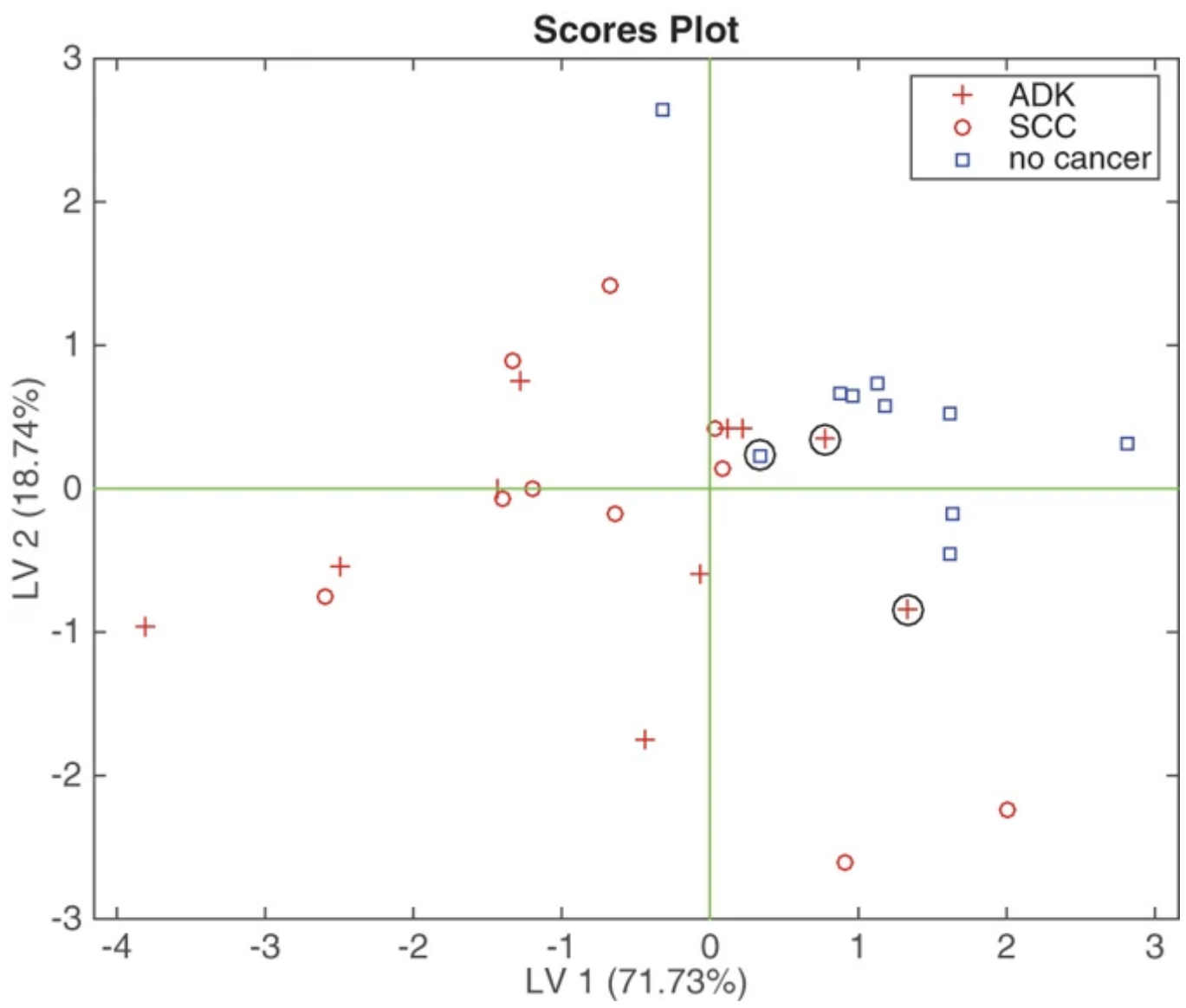

**Figure 2 . The plot of the first two latent variables of the PLS-DA model aimed at classifying cancer and non-cancer from the electronic nose data related to the mixed-expired breath.**

Only 56% of the cancer data were correctly classified between ADK and SCC when PLS-DA was used exclusively on those data. Because there is no distinction between the two, it is likely that the concentration of VOCs needed to tell ADK from SCC is lower than the resolution of the sensors. To the best of our knowledge, however, differing VOC patterns between ADK and SCC have been seen in histology samples.

It was interesting to discover that the categorization outcomes were unrelated to the lung from which the samples were taken. As with samples taken from inside the lungs, the proportion of accurate classifications obtained with mixed-expired air was comparable. Air taken from the opposing lung somewhat increased the rate of accurate classifications. The classifiers worked well, correctly classifying damaged lungs 93% of the time and unaffected lungs 96% of the time.

Figures 3a and 3b, respectively, provide the score plots for the first two latent variables using samples taken from the ipsilateral and contralateral lungs. It is noteworthy that mixed-expired breath and samples taken from the lungs both performed equally well in detecting lung cancer when using the identical porphyrins-coated QMB array.

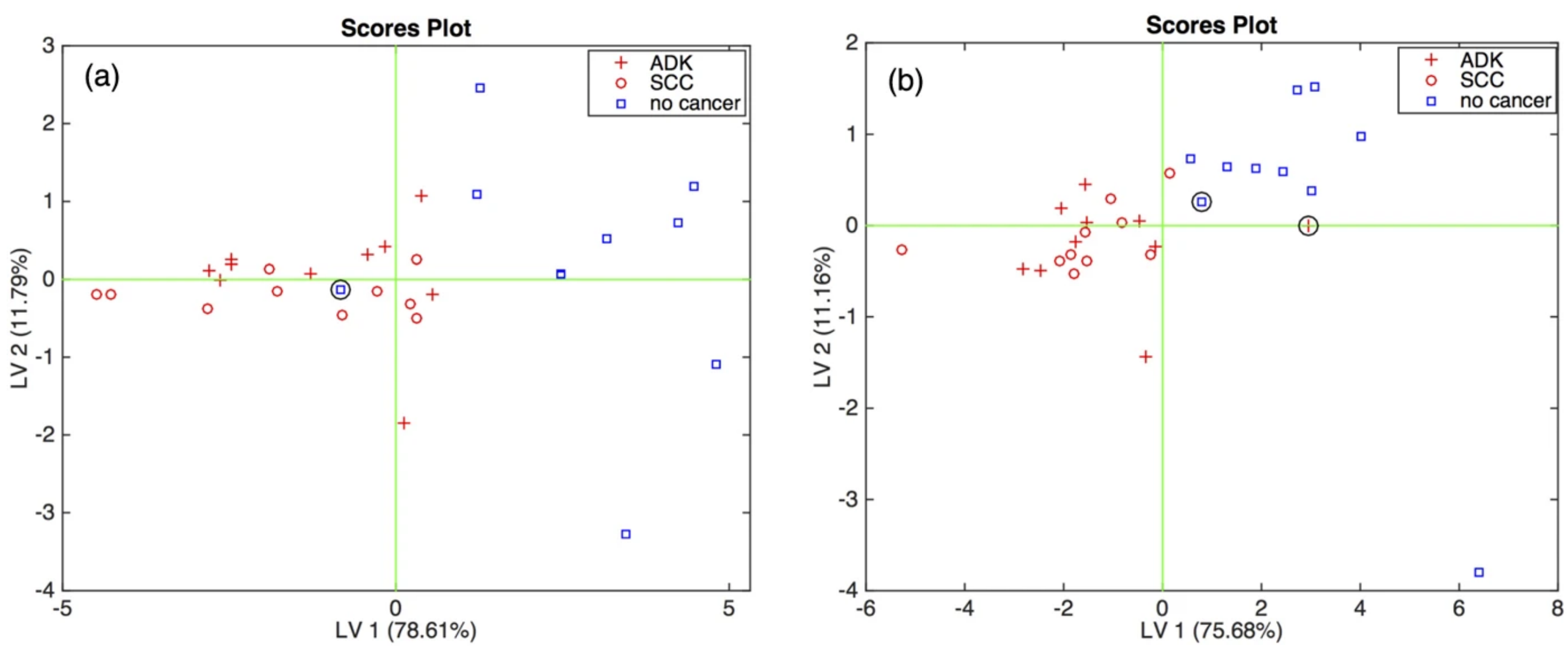

**Figure 3. Classifies cancer and non-cancer from the electronic nose related to the air sampled from inside the affected lung (Fig. 3a) and not-affected lung (Fig. 3b).**

The signals of some of the sensors on the array are consistent with the class to which the sample belongs, while other sensors on the array provide signals that differ from the expected average for that class, according to a comparison of the sensor patterns of misclassified samples with the mean pattern of each class. The sensor patterns of the misclassified samples indicate that the misclassifications may be caused by changes in specific substances to which some sensors are more sensitive, even if a thorough analysis of the link between individual sensor signals and cancer is outside the scope of this research. This suggests that these categorization mistakes may be completely eliminated by a sensor array with good design.

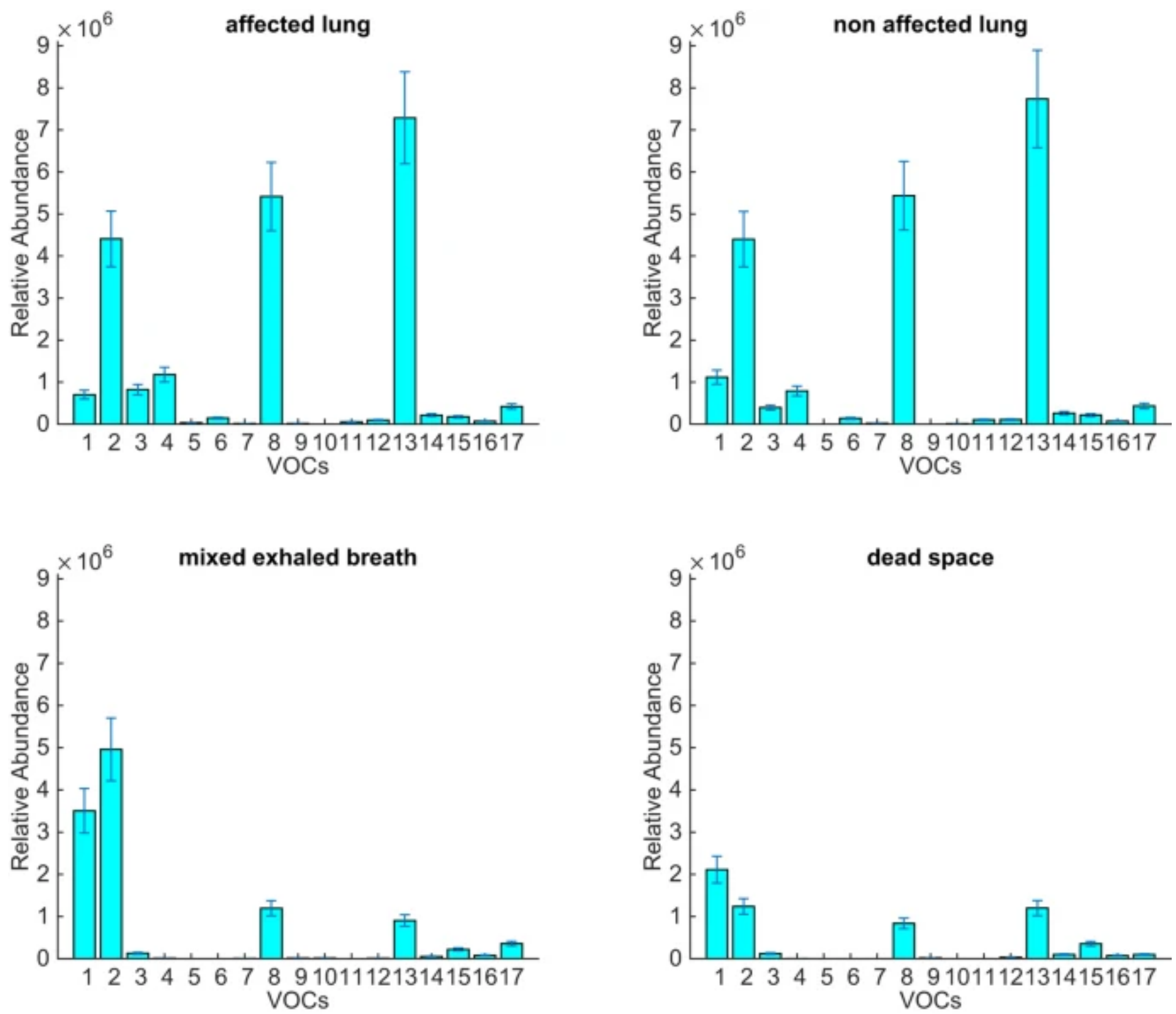

**Figure 4. Average abundance of the VOCs found in the four kinds of measured samples.**

The plot's findings show that the profiles of the two in-situ collected samples are noticeably comparable. The mixed-expired breath and the dead space both included many of the same substances that were in the lungs. All chemicals showed a general decline in quantity, with the exception of ethanol, whose abundance was higher in the lung than the breath. This suggests that the upper respiratory system produces a significant amount of this substance, and ethanol is proposed as a possible marker for head and neck cancers.

These results support the electronic nose findings and imply that certain parts of the VOCs pattern are preserved by the breath collecting technique. It should be observed that certain molecules, especially 8 and 12, are less abundant in the mixed-expired breath than they are in the lungs, suggesting that the pattern of VOCs that originates in the lungs may change as it passes through the airways.These adjustments have no impact on the electronic nose's ability to distinguish between cancer and other ailments. It would be interesting to find out if GC-MS can identify any variations between the two lungs of the same patient. Each VOC's p-value from the Kruskal-Wallis rank test is listed, and it was discovered that there was no significant difference between any particular VOCs in the two lungs.

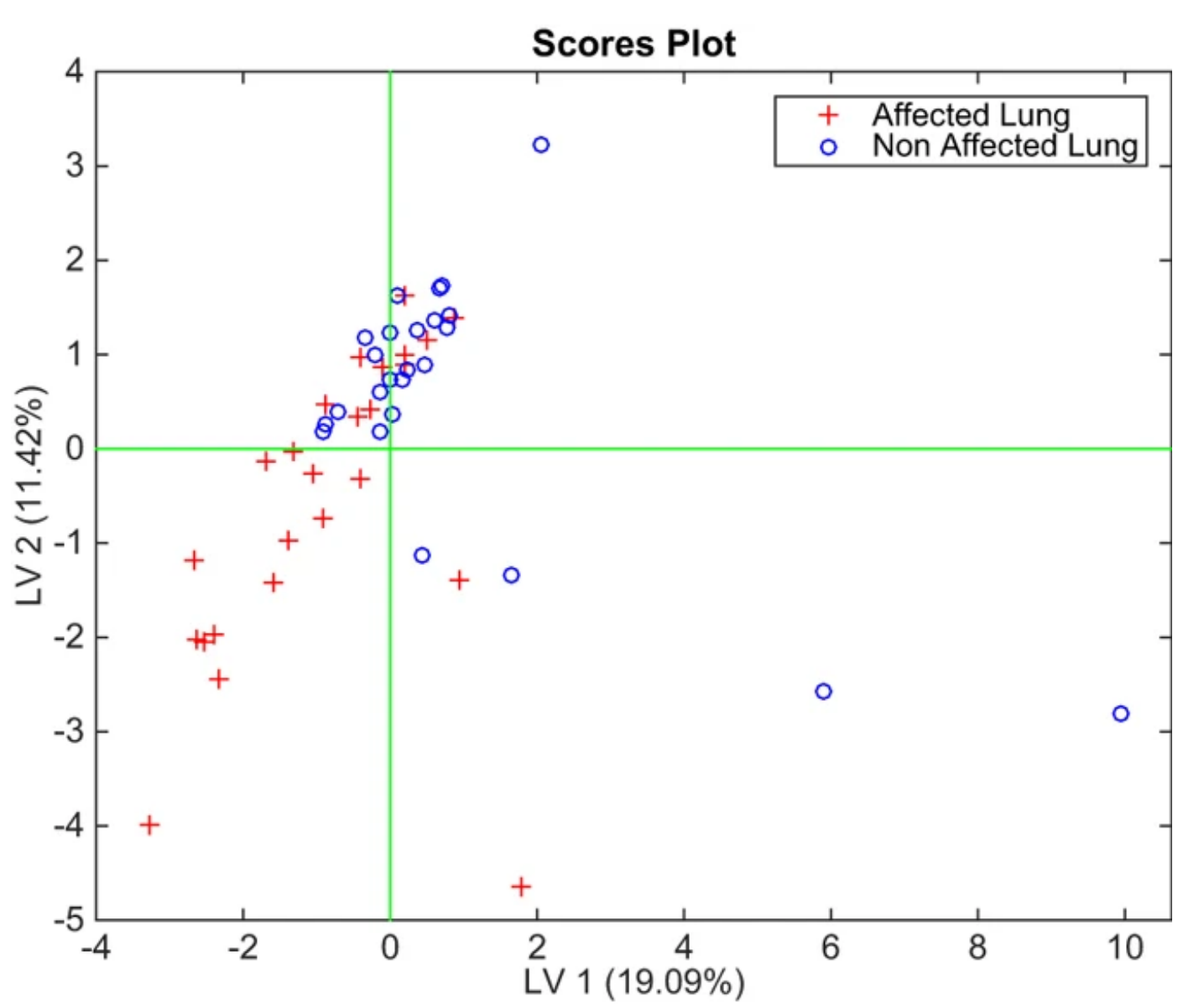

**Figure 5. The abundance of 20 different air-sampled compounds**

These findings imply that there are some differences between the lungs and that certain VOCs are expelled directly from cells. However, this number is too little to be picked up by the sensors in the electronic nose or identified by GC-MS. These findings agree with those obtained from VOC sampling performed after surgical tumor excision employing one-lung breathing on both lungs. In that experiment, PLS-DA completely distinguished between the lungs prior to tumor excision. The identical SPME and GC-MS apparatus was utilized by Wang et al.20 in their investigation, although in this study, samples were taken from patients who were in different states during a bronchoscopic examination.

## Discussion and Conclusion

The kind of lung from which the sample is taken does not significantly affect the electronic nose's capacity to detect lung cancer. This could be as a result of the experiment's sensors' lack of sensitivity to specific essential molecules present in the injured lung. It is significant to remember that these results are unique to the electronic nose utilized in this investigation. Even though only 76% of the two lungs were correctly identified, GC-MS research reveals that there may be minute changes between the two lungs.

This investigation also focused on the passage of VOCs from the lungs via the airways. With the exception of ethanol, which is probably formed in the upper airways, VOCs are mainly retained in breath (dead space and mixed-expired breath) although their relative abundance is more diluted there. Electronic nose sensors rely on a pattern, however the dilution factor is not constant for all chemicals. When measuring air from the lungs and breath, cancer and non-cancer samples may be distinguished with equal effectiveness. However, it cannot be completely disregarded because these variations may be significant when categorizing for other ailments. To better understand the spread of VOCs and how to improve monitoring techniques, more study is required.

It's crucial to keep in mind that the sample procedure entails gathering breath in bags at room temperature. Used in this investigation were tedlar bags. Tedlar is a widely used material for collecting breath because of its outstanding background emission and VOC stability qualities. However, condensation can occur when there is a temperature difference between the air inside the body and the outside air, which lowers the concentration of less volatile chemicals. Furthermore, since the bag material's inertness cannot be completely disregarded, it is possible that some substances may adhere to the surface. This study confirms that breath analysis is useful in identifying lung cancer, but it also raises the potential that cancer-related VOCs may not be specific for a certain type of cancer but rather indicative of malignancies that manifest in other body parts.

The samples were split up differently between the GC-MS and the electronic nose. The electronic nose was utilized to measure all samples with the exception of the dead space. Only for the subset of cancer patients were all four types of samples evaluated using GC-MS.

## Statement of Outside Assistance

JSHS
JUNIOR SCIENCE AND HUMANITIES SYMPOSIUM

### National Junior Science & Humanities Symposium
### Statement of Outside Assistance

Students submitting their research paper to the Regional and National symposium must complete this form in full and submit with the final research paper. Please type "N/A" in any field that is "not applicable" to your research.

**Student / Participant to Complete**

**Name:** Jayanth Mokkapati

**Regional Symposium:** Dallas

**Title of Paper:** Lung Cancer Diagnosing Biomarker Through Breath Samples for Respiratory Evaluation

1. **Please explain your role in the development of the project idea.**

Steps completed in development of machine:

The mixed-expired breath and air taken from inside the lungs were measured using an electronic nose. In every case, electronic nose signals were utilized to differentiate between cancer and non-cancer patients.

With a three latent variables model, the Partial Least Squares Discriminant Analysis (PLS-DA) applied to electronic nasal data connected to mixed-expired breath resulted in a 93 percent accurate classification rate. A leave-one-out technique was used to identify the number of latent variables, with the goal of reducing prediction error. The PLS-DA models mentioned here have not been tested by an independent data set due to the esiguity of the data set. Nonetheless, cross-validated PLS-DA has been shown to be equal to a multivariate ANOVA statistical test, with the added feature of a scores plot that allows the reciprocal connection between the data to be visually recognized.

The electronic nose data clustering is seen in the figure 1, which displays the plot of the first two latent variables. This conclusion is consistent with prior discoveries employing the same sensor technology in electronic noses. The propensity of cancer data to be more dispersed than controls is verified, especially when the identical breath collecting approach was used.

2. **What steps led you to formulate your research question?**
   **– or – What steps led you to develop the design for your project?**

Determine early stages of lung cancer to prevent development through breath samples
Create a very accurate machine that can determine percent composition of different compounds
Create a fast and efficient machine to use
Create a Low-Cost and very affordable machine for an average consumer
Product shall be very user friendly that can be used with ease

3. **Where did you conduct the major part of your work?**
   *(e.g., home, school, or other institutional setting – university lab, medical center, etc.)*

Home

Junior Science and Humanities Symposium
jshs.org | 1-703-312-9206 | admin@jshs.org

**The Student, Teacher and/or Supervising Mentor must sign below.** ***If you did the work without a teacher or supervising scientist, you will need a signature from your parent and a brief description of their role in the research.***

| | | |
|---|---|---|
| 12-15-22 | JM- Signature | Plano West |
| Date | Signature of **Student** (required) | **Student's** High School |
| | | |
| Date | Signature of **Teacher** | **Teacher's** High School |
| | | |
| Date | Signature of **Supervising Mentor** | Name of **Supervising Mentor** |
| | | |
| | Title of **Supervising Mentor** | Institution of **Supervising Mentor** |
| 12-15-22 | SM- Signature | Swapna Mokkapati 2108356263 |
| Date | Signature of **Parent** (required if no Teacher or Supervising Mentor was involved in the research.) | Name and Phone Number of **Parent** |

JSHS Statement of Outside Assistance v2021 4

4. **Describe the assistance that you received throughout the project.**
   I recieved minimal assistance from my parents while dealing while eletronic circuits.

5. **If you worked in an institutional setting, describe your role on the team.**
   NA

6. **What role did each person play in the research investigation?**
   NA

7. **Describe what parts of the research you did on your own and what parts where you received help.** *(e.g., literature search, hypothesis, experimental design, use of special equipment, gathering data, evaluation of data, statistical analysis, conclusions, and preparation of written report (abstract and/or paper))*

In recent years, electronic nose devices have become a popular approach for identifying respiratory disorders including lung cancer. Traditional e-nose systems have had very consistent principles and patterns of sensor responses. After coming to the realization that detecting cancer at early stages can save 99 percent of lives, it has become imperative to design a machine that can easily detect for lung cancer(the most common type of cancer) in a cost-effective and accurate way. Designing an AI Nose was a perfect way to counteract the problem.

A tiny e-nose system with 14 gas sensors of four types was created and fifty breath samples were analyzed. Five feature extraction techniques and two classifiers were used to test the system's efficiency in recognizing and discriminating lung cancer from other respiratory disorders and healthy controls. Finally, the impact of different sensor types on the capacity of e-nose systems to identify objects was investigated. The sensitivity, specificity, and accuracy of distinguishing lung cancer patients from healthy controls are 91.58 percent, 91.72 percent, and 91.59 percent, respectively, when utilizing the DA fuzzy 5-NIN classification approach.

The findings imply that type-specific sensors might greatly improve the diagnostic accuracy of e-nose devices. These findings indicated that the e-nose system described in this work might be used in lung cancer screening with good results. Furthermore, while creating e-nose systems, it is critical to consider type-different sensors. This machine covers all aspects to most effectively develop a machine that can detect lung cancer that is user-friendly and low cost.

JSHS Statement of Outside Assistance v2021 2

8. **If this research is a continuation of an investigation that was previously submitted to a regional JSHS, describe how you have expanded your investigation.**
   NA

**Teacher and / or Supervising Mentor to Complete (or Parent if no teacher / mentor involvement):**
**Comments by teacher and/or supervising mentor on the students' individual contributions to the research investigation or engineering/computer science project.** *If no Teacher or Mentor/Scientist was involved the Parent must complete this section describing their role in the research.*

The role of the parent was to make sure the student was taking proper safety precautions when dealing with eletronic circuits.

**Statement by the teacher and/or supervising mentor acknowledging that the student conducted the research in accordance with proper procedures and protocols for the conduct of animal research or human research.**
*Projects which were conducted without proper supervision will be disqualified from both regional and National competition. Further guidelines may be found at http://www.jshs.org*

- Research activities involving **non-human vertebrates** or **human subjects** must be submitted for IRB review prior to the conduct of the research.
- Research activities **involving vertebrate animals** must be conducted in compliance with local, state, and federal guidelines for the humane and ethical treatment of animals in the conduct of the research.

JSHS Statement of Outside Assistance v2021 3